\documentclass[double]{article}
\usepackage{times}
\usepackage{algorithm}
\usepackage{algorithmic}
\usepackage{wrapfig}
\usepackage{verbatim}
\usepackage{amsmath}
\usepackage{graphicx}
\usepackage{subcaption}
\usepackage[active]{srcltx}
\usepackage{color}
\usepackage{lineno}
\usepackage{dirtytalk}
\usepackage{amsthm}
\usepackage{authblk}
\usepackage{listings}
\usepackage{setspace} 
\usepackage{tikz}
\usetikzlibrary{shapes.geometric, arrows}

\usepackage{epsfig,graphicx,amsfonts}
\usepackage{amsmath,amsthm,amssymb}
\usepackage{xcolor}

\usepackage{adjustbox}
\usepackage{multirow}
\usepackage{setspace}
\usepackage{hyperref}
\usepackage{comment}

\long\def\comment #1\commentend{}

\begin{document}
\doublespacing

\title{\Large The Topology of a Family Tree Graph and Its Members' Satisfaction with One Another: A Machine Learning Approach}
\author{Teddy Lazebnik$^{1,2,*}$, Amit Yaniv-Rosenfeld$^{3,4,5}$ \\ \(^1\) Department Mathematical, Ariel University, Ariel, Israel\\ \(^2\) Department of Cancer Biology, Cancer Institute, University College London, London, UK \\ \(^3\) Shalvata Mental Health Care Center, Hod Hasharon, Israel\\ \(^4\) Sacklar Faculty of Medicine, Tel-Aviv University, Tel-Aviv, Israel\\ \(^5\)Department of Management, Bar-Ilan University, Ramat-Gan, Israel\\ \(*\) Corresponding author: lazebnik.teddy@gmail.com }

\date{}

\maketitle 

\begin{abstract}

Family members' satisfaction with one another is central to creating healthy and supportive family environments. In this work, we propose and implement a novel computational technique aimed at exploring the possible relationship between the topology of a given family tree graph and its members' satisfaction with one another. Through an extensive empirical evaluation ($N=486$ families), we show that the proposed technique brings about highly accurate results in predicting family members' satisfaction with one another based solely on the family graph's topology. Furthermore, the results indicate that our technique favorably compares to baseline regression models which rely on established features associated with family members' satisfaction with one another in prior literature. \\

\end{abstract}

\maketitle \thispagestyle{empty}

\pagestyle{myheadings} \markboth{Draft:  \today}{Draft:  \today}
\setcounter{page}{1}

\textit{\say{All happy families are alike; each unhappy family is unhappy in its own way.}} Anna Karenina (Leo Tolstoy). 

\section{Introduction}
\label{sec:introduction}
Family dynamics and relationships are fundamental to human life, influencing individual well-being, social development, and emotional health \cite{intro_2,intro_3,intro_4,intro_5}. In fact, understanding the factors associated with family members' satisfaction with one another is a focal research challenge inherent to various fields including sociology \cite{intro_6,intro_7}, mental healthcare \cite{intro_13,intro_14}, psychology \cite{intro_12,intro_16}, and family studies \cite{intro_8,intro_9,intro_10}, to name a few. Broadly speaking, research efforts in this realm seek to uncover the underlying mechanisms of human social interactions and emotional bonds within the family social unit to extend our psycho-social understanding of family dynamics and to provide a foundation for developing strategies and interventions aimed at enhancing family well-being and fostering healthier, more supportive family environments \cite{new_intro_1,new_intro_2,new_intro_3}. 

One classic technique to represent a family is by using a tree graph structure \cite{family_tree_def}. In such a tree graph, nodes represent the family members (possibly along with some basic information about each of them such as name, gender, and birth and/or demise dates) and edges represent relationships native to family dynamics such as marriage, divorce, and parenthood. Family tree graphs may be highly complex \cite{ft_1,ft_2} and may vary widely in their topology from one family to the other due to cultural, demographic, and economic factors, to name a few \cite{family_tree_def}. From a research methodology perspective, in order to overcome these challenges, prior literature has typically focused on a limited number of aspects within the family tree graph and investigated their possible relationships with family members' satisfaction with one another. For example, extensive lines of work have focused almost exclusively on parent-child characteristics \cite{end_intro_1}, sibling relationships \cite{end_intro_2}, marital dynamics \cite{end_intro_3}, and other specific aspects of the family tree graph. That is, to the best of our understanding, the \textit{topology} of a family tree graph, i.e., its entire structure, is not considered as the basic unit of analysis in existing literature. 

In this work, we propose and implement a novel computational technique aimed at exploring the relationship between the topology of a family tree graph and its members' satisfaction with each other.
In particular, our technique is based on a machine-learning pipeline consisting of a deep-learning AutoEncoder and a supervised regression model. Initially, the family tree graph is converted into informative feature vectors using a variational graph AutoEncoder model \cite{graph_autoencoder}. In turn, these feature vectors are used for training supervised learning regression models to estimate specific measures of family members' satisfaction with one another such as the average satisfaction in the entire tree graph (average satisfaction between any two family members) and/or the average satisfaction of a given member with his/her nuclear family members (for married and/or parents: spouse and children; otherwise: with parents and siblings) given the family tree graph's topology. Using real-world data from 486 families, we show that our technique is highly successful in predicting family members' satisfaction with one another based \textit{solely} on the family graph's topology. Furthermore, we show that our technique favorably compares to baseline regression models which rely on established features associated with family members' satisfaction with one another in prior literature. 

The rest of the article is organized as follows: In Section \ref{sec:RW}, we review relevant prior work. Then, in Section \ref{sec:methods}, we formally present the proposed technique. In Section \ref{sec:empirical}, we apply and compare our technique with two baseline models using family tree graphs and satisfaction scores we curated. Finally, in Section \ref{sec:discussion}, we discuss and interpret the results, highlighting possible implications and future work directions. 

\section{Related Work}\label{sec:RW}

A family tree graph is a popular representation of relationships between individuals or even groups where one emerges from another \cite{ft_data_stracture}. Its use is not restricted to the social sciences alone \cite{ft_intro,sturgess2001young,foxman1989family} and, in fact, it offers an imperative tool in various other fields ranging from biology, where family tree graphs are common for representing and studying evolution processes \cite{biology_tree}, to public health where historical records of family members in the family tree are used to infer disease probabilities in individuals \cite{new_rw_1}.

Focusing on the use of family tree graphs in the social sciences, prior studies have explored the possible connection between family tree graph topologies and various socio-economic phenomena such as economic stability, educational attainment, and health. 
For instance, \cite{ft_indicator_example} reviewed studies of income inequality and family structure changes and found a strong correlation between family structure (i.e., topology) and class, race, and gender inequalities. In a similar manner, \cite{ft_indicator_example_2} used data from the Michigan Panel Study of Income Dynamics (MPSID)  to explore the relationship between female-headed families and economic deprivation. \cite{new_rw_2} examined the effects of family structure on educational attainment after controlling for common family influences, observed and unobserved, using data from siblings. The authors found that family structure is a statistically significant indicator of educational attainment. \cite{new_rw_3} invstigated the relationship between family structure and children's access to health care, in general, and the extent to which family structure types predict children's utilization of preventive health care, and barriers to care, in particular. The authors found that children of single mothers demonstrate extremely different patterns of healthcare access than do the children of single fathers. 

Focusing on the study of family members' satisfaction with one another, a large body of work has considered specific aspects of the family tree graph \cite{new_rw_4,new_rw_6,new_rw_5}. For instance, \cite{georgas2001functional} investigated the relationship between the family structure and other cultural and functional aspects of the family such as emotional distance, social interaction, and communication quality. As mentioned before, family tree graphs may be highly complex and vary widely in their topology from one family to the other \cite{family_tree_def}. As a result, these and similar prior studies often examine \say{direct} factors such as individual and relational attributes, rather than the overall structure of family tree graphs. Specifically, prior studies in the realm of family memebrs' satisfaction from one another have focused a wide range of specific aspects within the family tree graph, such as the \say{nuclear family} size \cite{p1_1,p1_2,p1_3,delpiano2006impacts}; male-to-female ratio \cite{p2_1,p2_2}; generation age difference \cite{p3_1,p3_2}; step-siblings and divorce \cite{p4_1}; the oldest generation members' average satisfaction \cite{p5_1}. 
To the best of our knowledge, this is the first work to consider the entire structure (i.e., topology) of the family tree graph as the basic unit of analysis. Nevertheless, the above-mentioned \say{direct} characteristics of the family tree graph are used in our analysis as well for the baseline modeling (formally outlined in Section \ref{sec:empirical}).   
 
In the context of family members' satisfaction with one another, the terms \say{nuclear family} and \say{extended family} commonly emerge to differentiate between the immediate, and typically presumed most influential, family environment, and the more extended family environment \cite{nuclear_extended_definition}. Unfortunately, there seems to be little consensus on clear and formal definitions for these terms and different researchers tend to adopt slightly different interpretations  \cite{nuclear_extended_definition_2}. In this work, we follow one standard approach where the nuclear family is defined as the immediate connections within one's family tree graph: parents, siblings, spouse, and children; whereas any other member is considered as part of the extended family \cite{saggers2005diversity,garcimartin2012defining}. 
The conceptual separation between nuclear and extended family is prominent in the study of families \cite{howells1962nuclear}. 
For example, \cite{alawad1992childhood} studied the relationship between the family structure and emotional and social development in the Sudanese capital, Khartoum. Their analysis revealed that children living with their nuclear families had more conduct, emotional, and sleep problems, poorer self-care, and were more likely to be overdependent compared to those living with their extended families. In particular, the author shows that the grandmother's involvement was the strongest predictor of normal social and emotional adjustment for young children. \cite{new_rw_7} investigate the changes in later-life families (families where most members are 65 years old or older), focusing on how family caregiving occurs within these families. The authors show multiple differences between the nuclear and extended families in this context. As such, in this work, we too consider the nuclear family and extended family satisfaction separately. 

\section{Computational Technique}\label{sec:methods}

Formally, each family is represented as a tuple \(F := (G, S)\) where \(G := (V, E)\) is the family tree graph where \(V\) is the graph's nodes (i.e., family members) and \(E \subseteq V \times V \times T\) is the set of relationships between the family members and \(\forall \tau \in T := \{married, divorced, parent\}\). In addition, we assume access to a (possibly incomplete) satisfaction matrix \(S \in \mathbb{R}^{|V| \times |V|}\) where \(s_{i,j} \in S\) is the subjective satisfaction of family member \(i \in V\) with family member \(j \in V\). For convenience, let us assume $s_{i,j}\in [1,\ldots,10]$. Clearly, matrix \(S\) may not be symmetrical as family members \(i\) and \(j\) need not necessarily have the same satisfaction from one another (i.e., $s_{i,j}\neq s_{j,i}$).

At the core of the proposed technique is a two-step machine-learning pipeline. First, we use a deep neural network model to convert the family tree graph (\(G\)) into a representative feature vector. Deep neural networks are currently considered the state-of-the-art computational tool for representing graphs in a manner that captures complex patterns and relationships, making them particularly suitable for analyzing intricate data structures like family tree graphs \cite{nn_graph_best_1,nn_graph_best_2}. We adopt the Variational Graph AutoEncoder (VGAE) architecture as a representative model \cite{graph_autoencoder} given its promising results in similar tasks with limited-sized datasets in the past \cite{vgae_good}. The VGAE encodes a given family tree graph as a 16-dimensional feature vector. The resulting feature vectors are considered training features for a supervised regression model. Each of these vectors is then extended with some family satisfaction measure of interest as a target variable. 
Aligned with prior literature, we consider two common measures of family satisfaction: First, the Extended Family Satisfaction (EFS) measure which captures the average satisfaction between \textit{any} two members in the family tree graph (\(i,j \in V\)). Mathematically, EFS is defined as follows:
\begin{equation}
    EFS(F) := \frac{1}{|V|^2 - |V|}\sum_{i \neq j \in V} s_{i,j}
    \label{eq:efs}
\end{equation}
Second, we consider the Nuclear Family Satisfaction (NFS) measure which, given a specific family member \(i\in V\), captures the average satisfaction level between the members of the \say{immediate family}, denoted by \(\mu\). If the member is married and/or has children, \(\mu\) is defined as his/her spouse and children. Otherwise, \(\mu\) is defined to be one's parents and siblings. Formally, we define NFS as follows:
\begin{equation}
    NFS(F, i) := \frac{1}{|\mu|}\sum_{j \in \mu} s_{i,j}
    \label{eq:efs}
\end{equation}
Specifically, in our implementation, the family tree graph feature vectors are considered as training input for a supervised regression model with the EFS measure and the NFS measures associated with each member as target variables. 

\begin{figure}
    \centering
    \includegraphics[width=0.99\textwidth]{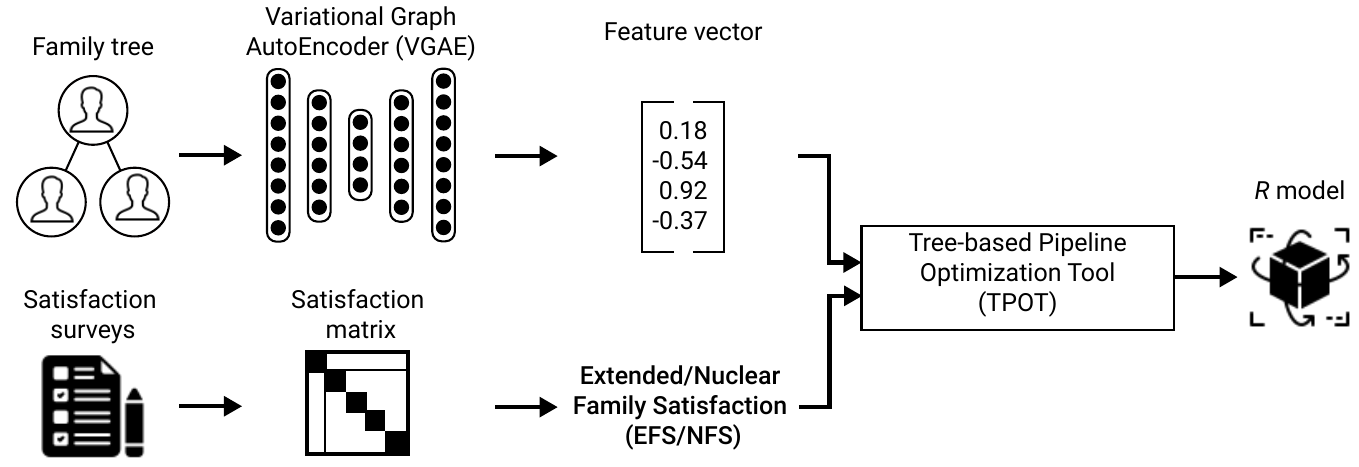}
    \caption{A schematic view of the proposed technique. The family tree graph is first converted by the VGAE model into a representative feature vector. In parallel, the satisfaction matrix is used for computing the family satisfaction measures of interest (e.g., the EFS and NFS measures). The extended feature vectors, now including the target variables, are then used as input to an optimization tool aimed at finding an optional regression model \(R\).}
    \label{fig:r_model}
\end{figure}

The resulting regression task is then solved using the Tree-based Pipeline Optimization Tool (TPOT) \cite{tpot}, an automated tool for optimizing the learning process using a genetic algorithm approach \cite{ga_1,ga_2,ga_3}, resulting in a regression model \(R\). For completeness, we consider two versions for $R$, a linear regression model denoted $R_l$ \cite{lg} and a non-linear version denoted $R_{nl}$, both automatically obtained using TPOT. For the training of the VGAE model, we used the entire dataset. For the training of \(R\), we use a typical 10-fold cross-validation method  \cite{k_fold}. Figure \ref{fig:r_model} shows a schematic view of the proposed technique. The models' hyperparameters and their values are provided in the Appendix. 

It is important to note that the feature vector representing the family tree graph is typically not self-explanatory \cite{deep_good} and, as such, it does not facilitate a \say{direct} examination of any specific family tree property of interest due to its highly complex non-linear computation. Hence, in order to investigate specific aspects of interest in the family tree graph (e.g., nuclear family size, male-to-female ratios, generation age differences, step-siblings and divorce, oldest generation members' satisfaction, etc.), one needs to measure or compute them directly from the raw input \(F\).  

\section{Empirical Study}
\label{sec:empirical}
\begin{figure}[!ht]
     \centering
     \includegraphics[width=0.99\textwidth]{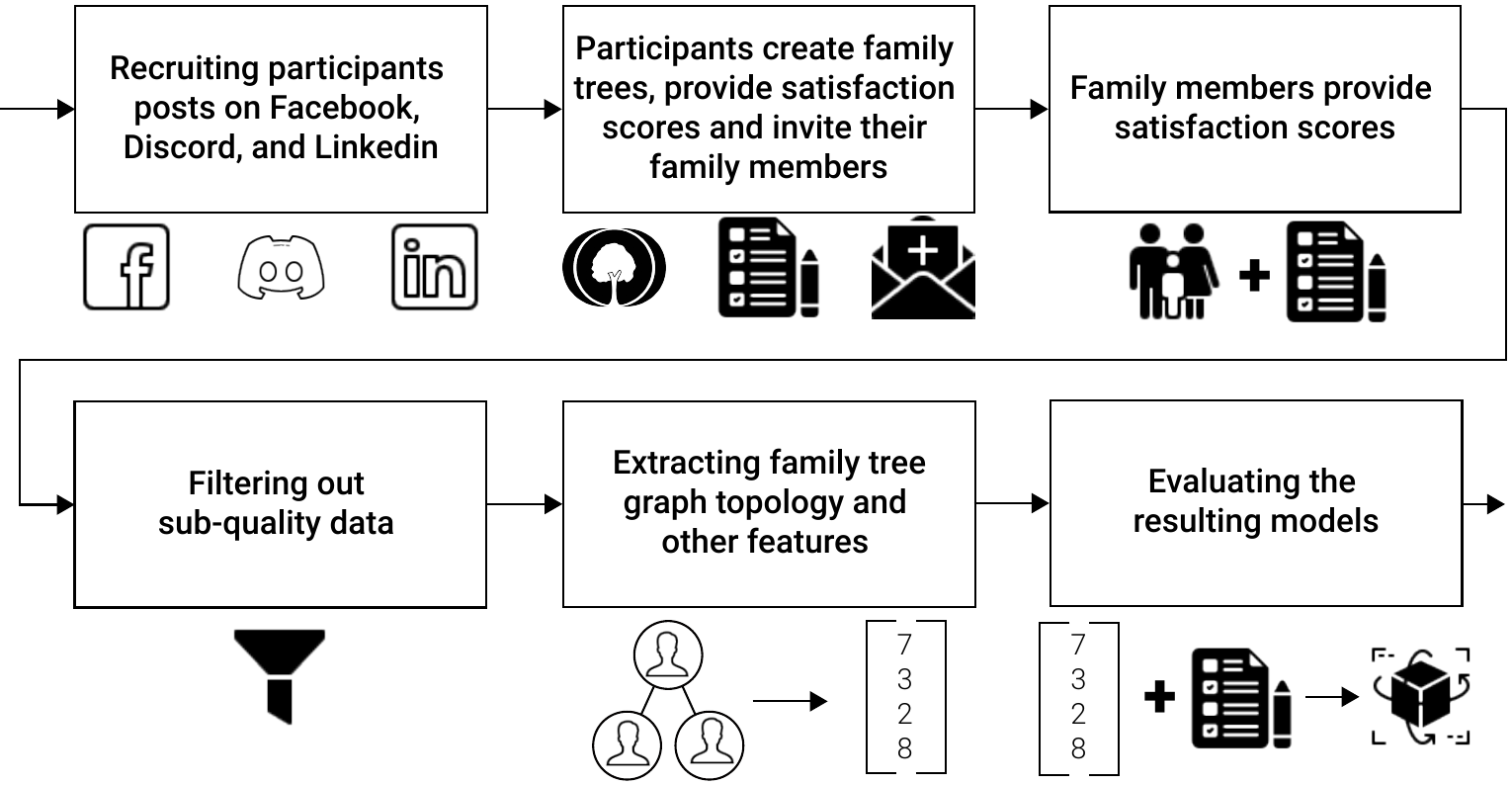}
     \caption{A schematic view of the empirical study's workflow.}
     \label{fig:flowchart}
 \end{figure}

In order to demonstrate the capabilities of our technique, we devise an empirical study: Initially, we collect a diverse dataset of family tree graphs and satisfaction scores. Then, we evaluate the proposed technique compared to baseline regression models using the $EFS$ and $NFS$ measures as target variables. Next, we detail the data collection process, followed by the implementation details of the baseline models, and conclude with the evaluation itself. Fig. \ref{fig:flowchart} presents a schematic view of the study's workflow. All procedures were approved by the corresponding Institutional Review Board (IRB). 

\subsection{Data Collection}
In order to obtain a diverse set of samples, we first posted invitations on several social networks including Facebook (\url{https://www.facebook.com/}), Discord (\url{https://discord.com/}), and LinkedIn (\url{https://linkedin.com/}). To minimize the possible bias of inviting participants from a specific social group (e.g., a single profession, association with some organization, or even country), we published multiple invitations on quasi-randomly chosen groups on social platforms that focused on hobbies such as football fans, fantasy books lovers, and home carpentry enthusiasts. 
Participants were not offered any compensation for their participation. 

Participants who answered our call, which we will refer to as  \say{seed participants}, digitally provided informed consent, had to declare that they are legally considered adults in their country, and are fluent in English. Seed participants were then asked to generate their digital family tree structure using the MyHeritage (\url{https://myheritage.com}) website which provides a simple and visual tool to generate family trees and export them. Participants were asked to place themselves at the center of the tree graph and detail at least two \say{vertical} generations (i.e., grandparents) and at least two \say{horizontal} generations  (i.e., second-order cousins, if exist). Then, each seed participant was asked to provide his/her subjective personal satisfaction from the relationship with each other family member in the tree using a 10-point Likert scale ranging from 1 (completely unsatisfied) to 10 (completely satisfied). After completing the above, the seed participant was asked to share a designated link (URL) with their family members who could, in turn, provide their own personal satisfaction scores with other family members of the same tree using the existing tree graph structure. It is important to note that family members could not observe the satisfaction scores provided by others and this fact was highlighted to all participants alike (both seed participants and those invited by them). 
All responses were kept confidential and anonymous as clearly stated to the participants at the beginning of the questionnaire. 

The data was collected between January 2023 and May 2023. Overall, about 93,000 individuals were exposed to our invitation to participate in this study. A sample of 617 individuals was obtained (roughly \(0.66\%\) response rate). To ensure the quality and completeness of our data, we restricted our analysis to families with at least three generations and a minimum of ten members (28 families omitted) and excluded families where less than 80\% of the adult family members participated in the study (10 families were omitted). In total, after accounting for possibly multiple seed participants from the same family, 486 families are considered in our subsequent analysis. 
These families are highly heterogeneous: considering the number of generations -- 91.77\% (446) of the families consists of three generations, 5.14\% (25) consists of four generations, and 3.09\% (15) consists of five generations. The number of family members also varied widely between 10 and 93 members with an average and standard deviation of \(32.47 \pm 8.04\). The EFS and NFS measures also present high variance with an average of 3.62-8.45 and 2.49-8.92 and a standard deviation of \(4.63 \pm 1.73\) and \(5.94 \pm 1.27\), respectively. 

\subsection{Baseline Models}

In order to fairly evaluate our technique, and specifically the $R_l$ and $R_{nl}$ models, we consider two additional baseline models that rely on features established in prior work as associated with family members' satisfaction with one another. Specifically, we consider a linear baseline model denoted $B_l$ and a non-linear baseline model denoted $B_{nl}$. These baseline models are trained using the same data as the \(R\) models but use a different data representation. Specifically, the baseline models use five features already established in the literature as linked with family members' satisfaction with one another - nuclear family size, male-to-female ratio, generation age differences, step-siblings and divorce, and oldest generation members' average satisfaction (as also discussed in Section \ref{sec:RW}). Formally, the nuclear family size is the count of nuclear family members divided by the size of the extended family (for normalization). The generation age difference is defined as the average difference between the average age across every two consecutive generations (indicated by the parent-children relationship). The step-siblings and divorce features are simple counts of the number of step-siblings and divorce in the entire family tree graph. The oldest generation members' average satisfaction is the average satisfaction score of the members of the most aged generation provided in the family tree graph (i.e., the first generation in the family tree graph if its members are vertically ordered using the parent-child relation).

\subsection{Results}
\label{sec:results}
\begin{figure}[!ht]
\centering
\begin{subfigure}{.99\linewidth}
\begin{subfigure}{.24\linewidth}
    \centering
\includegraphics[width=0.99\linewidth]{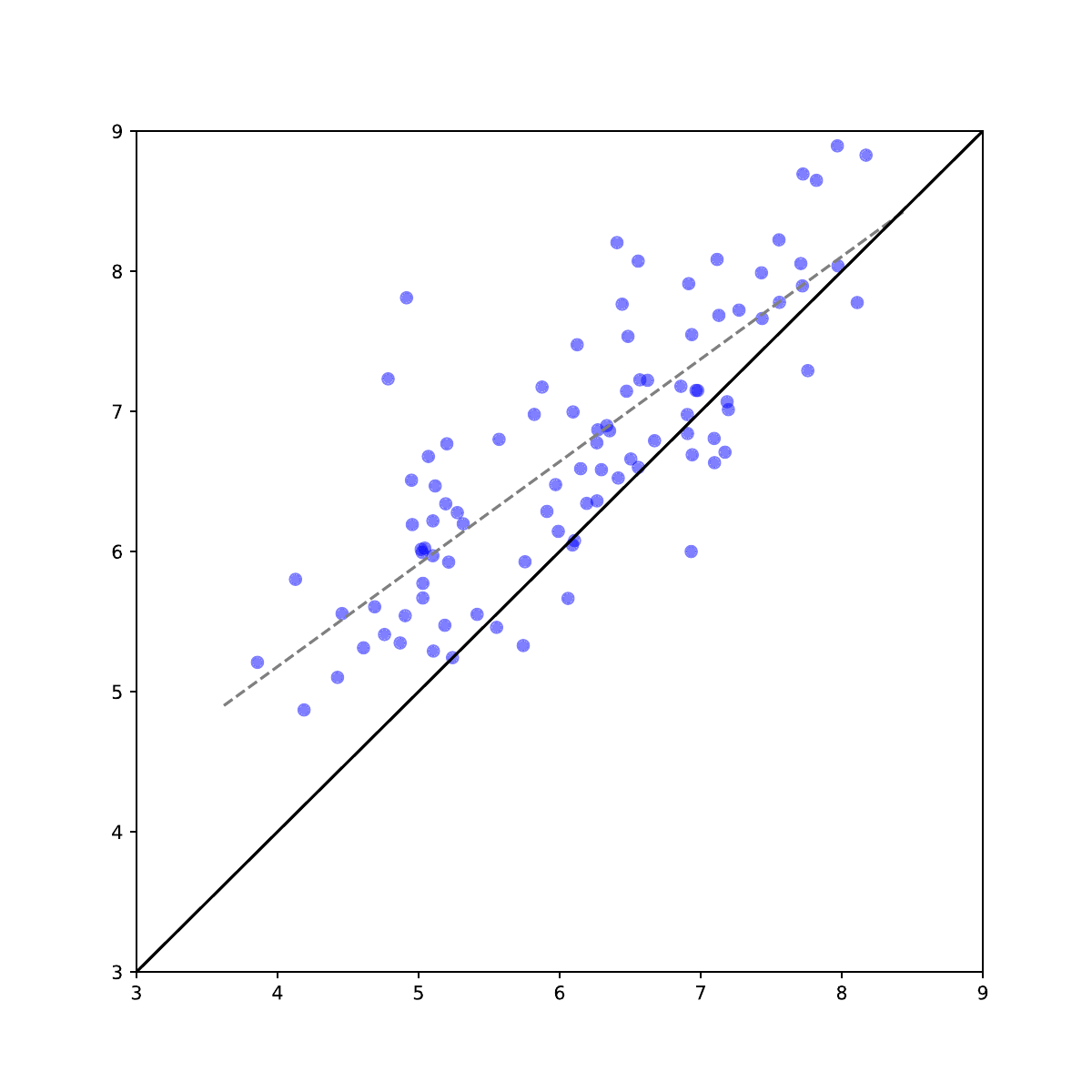}
    \caption{\(R_{l},0.7,0.63\).}
    \label{fig:3}
\end{subfigure}
\begin{subfigure}{.24\linewidth}
    \centering
\includegraphics[width=0.99\linewidth]{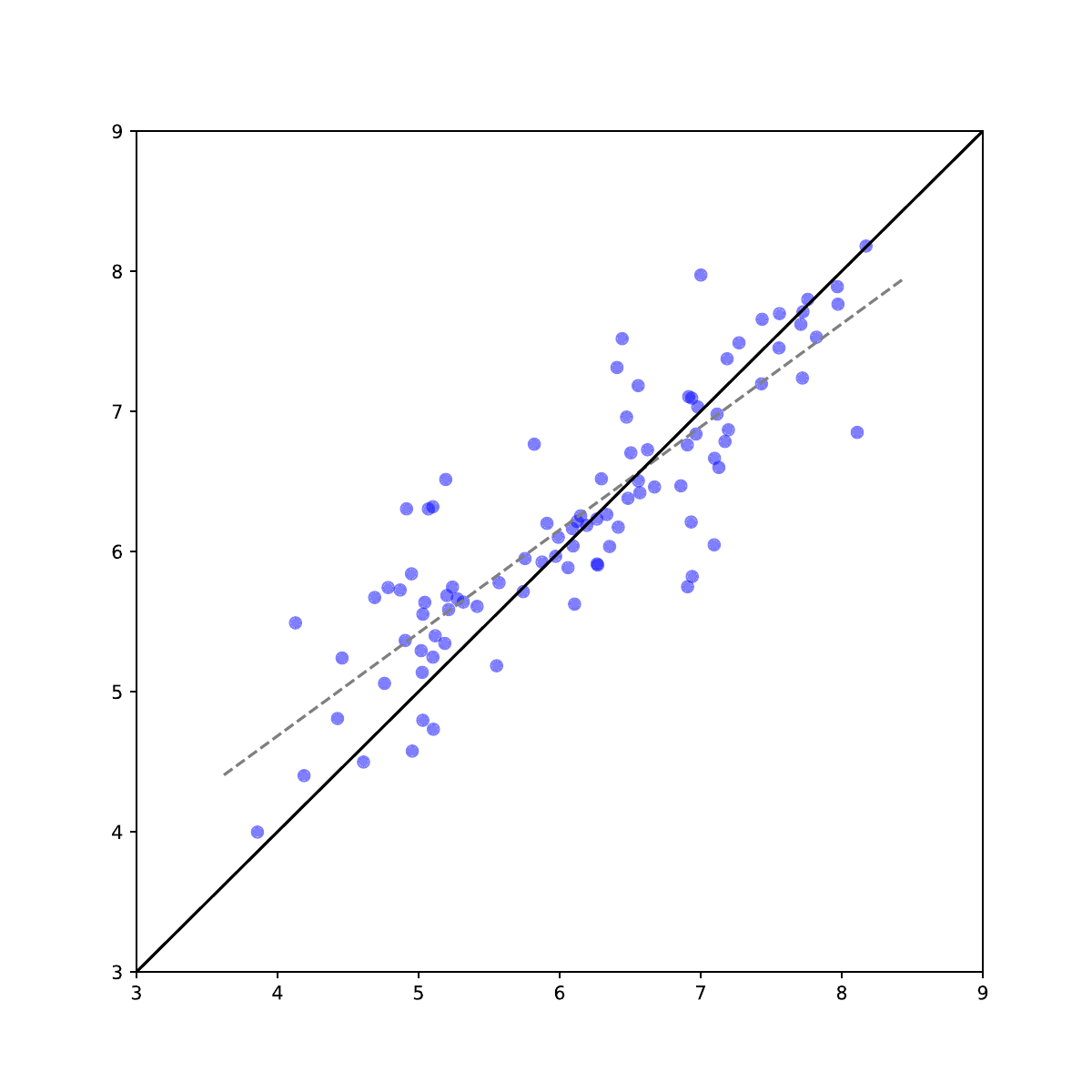}
    \caption{\(R_{nl},0.54,0.74\).}
    \label{fig:1}
\end{subfigure}
\begin{subfigure}{.24\linewidth}
    \centering
\includegraphics[width=0.99\linewidth]{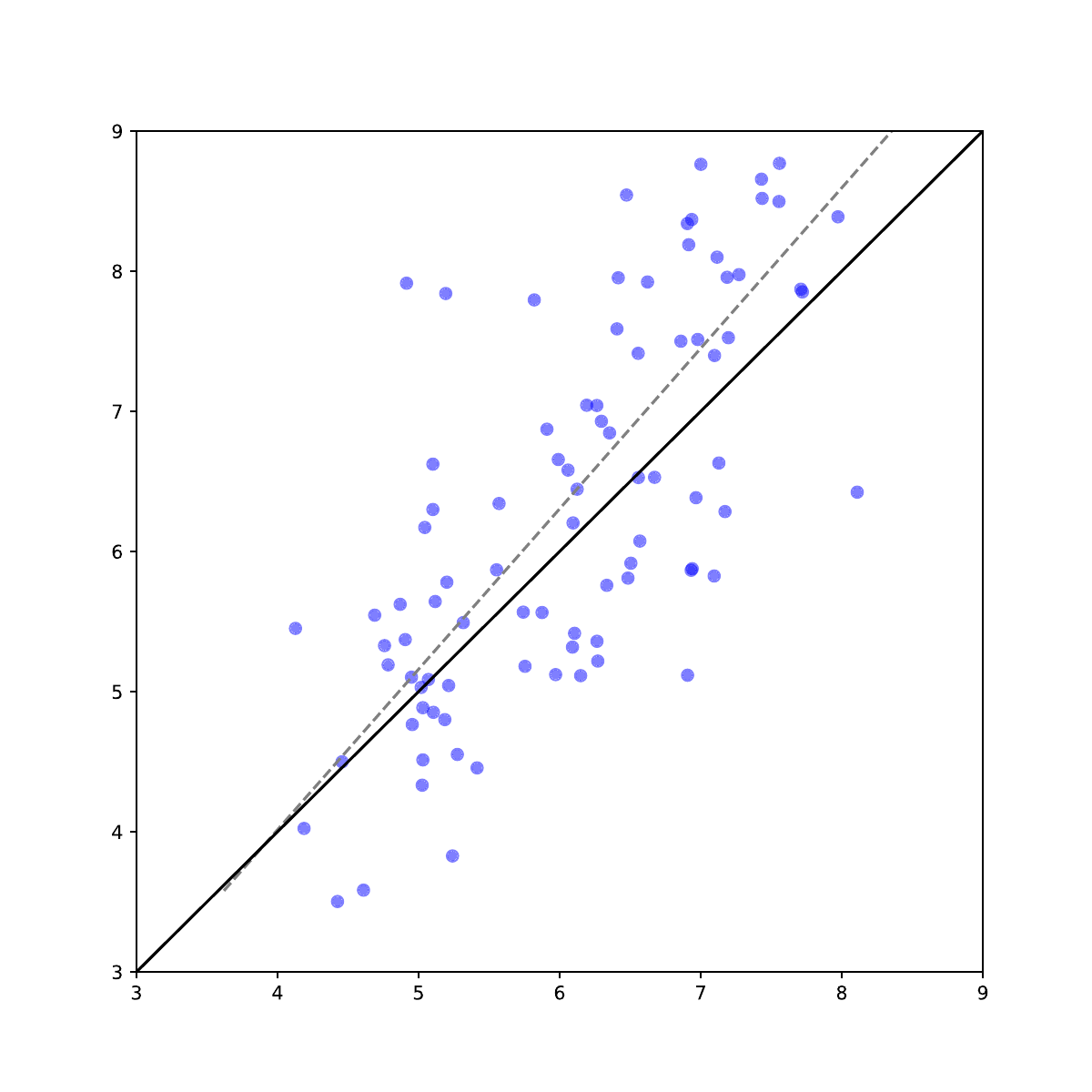}
    \caption{\(B_{nl}, 1.01, 0.58\). }
    \label{fig:7}
\end{subfigure}
\begin{subfigure}{.24\linewidth}
    \centering
\includegraphics[width=0.99\linewidth]{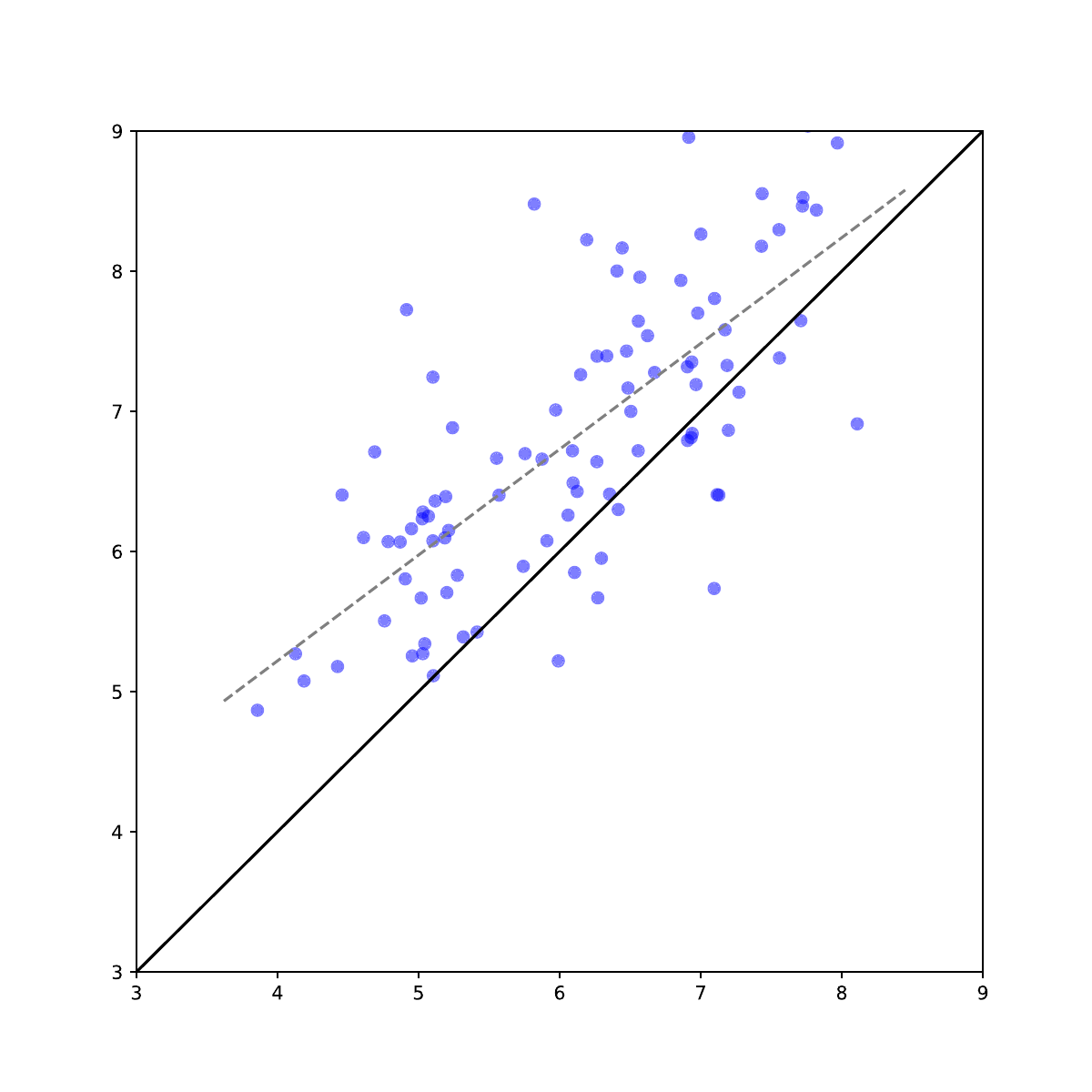}
    \caption{\(B_{l}, 0.84, 0.55\).}
    \label{fig:8}
\end{subfigure}\caption{Target Variable: EFS}
\end{subfigure}
\begin{subfigure}{.99\linewidth}
    \begin{subfigure}{.24\linewidth}
    \centering
\includegraphics[width=0.99\linewidth]{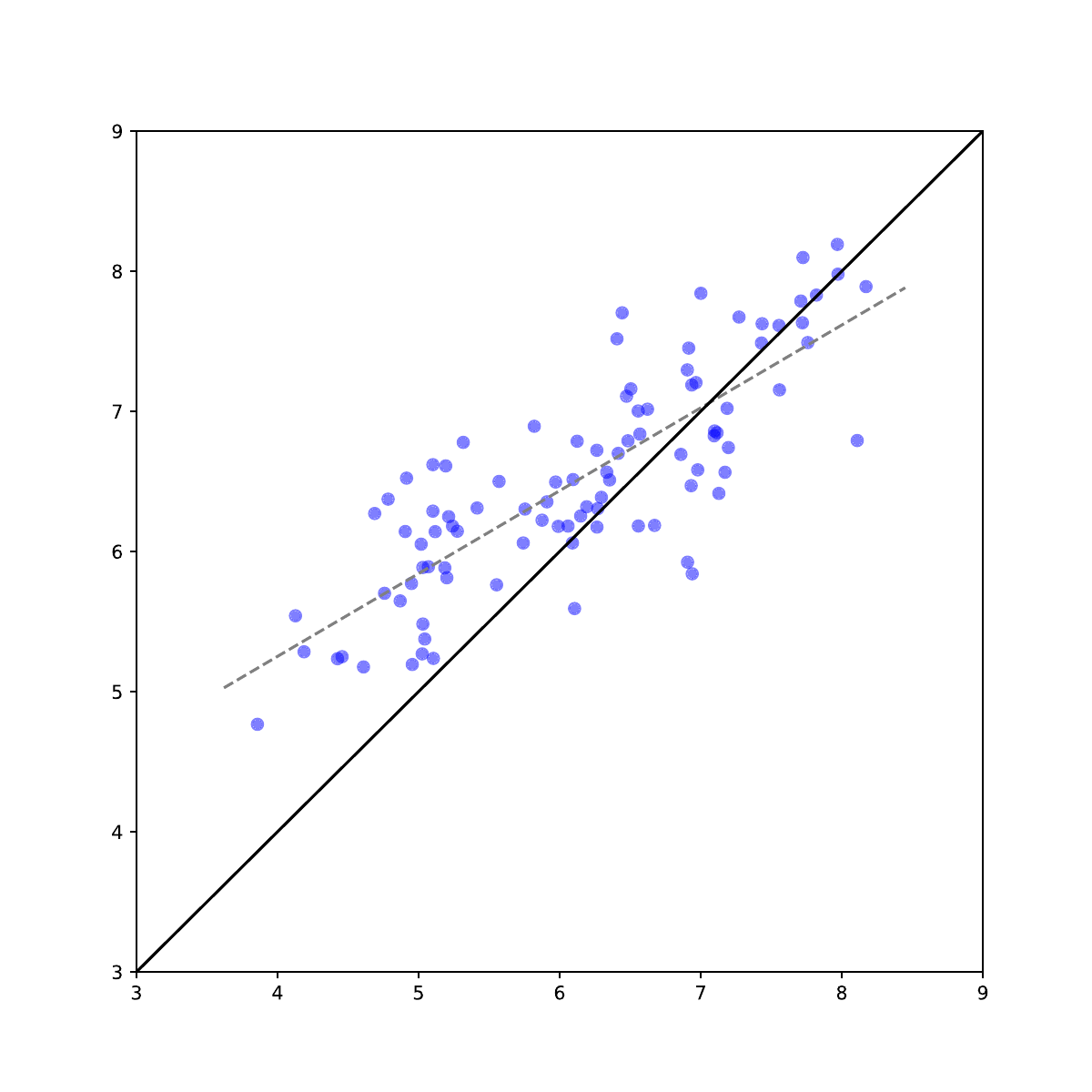}
    \caption{\(R_{l}, 0.58, 0.67\). }
    \label{fig:4}
\end{subfigure}
\begin{subfigure}{.24\linewidth}
    \centering
    \includegraphics[width=0.99\linewidth]{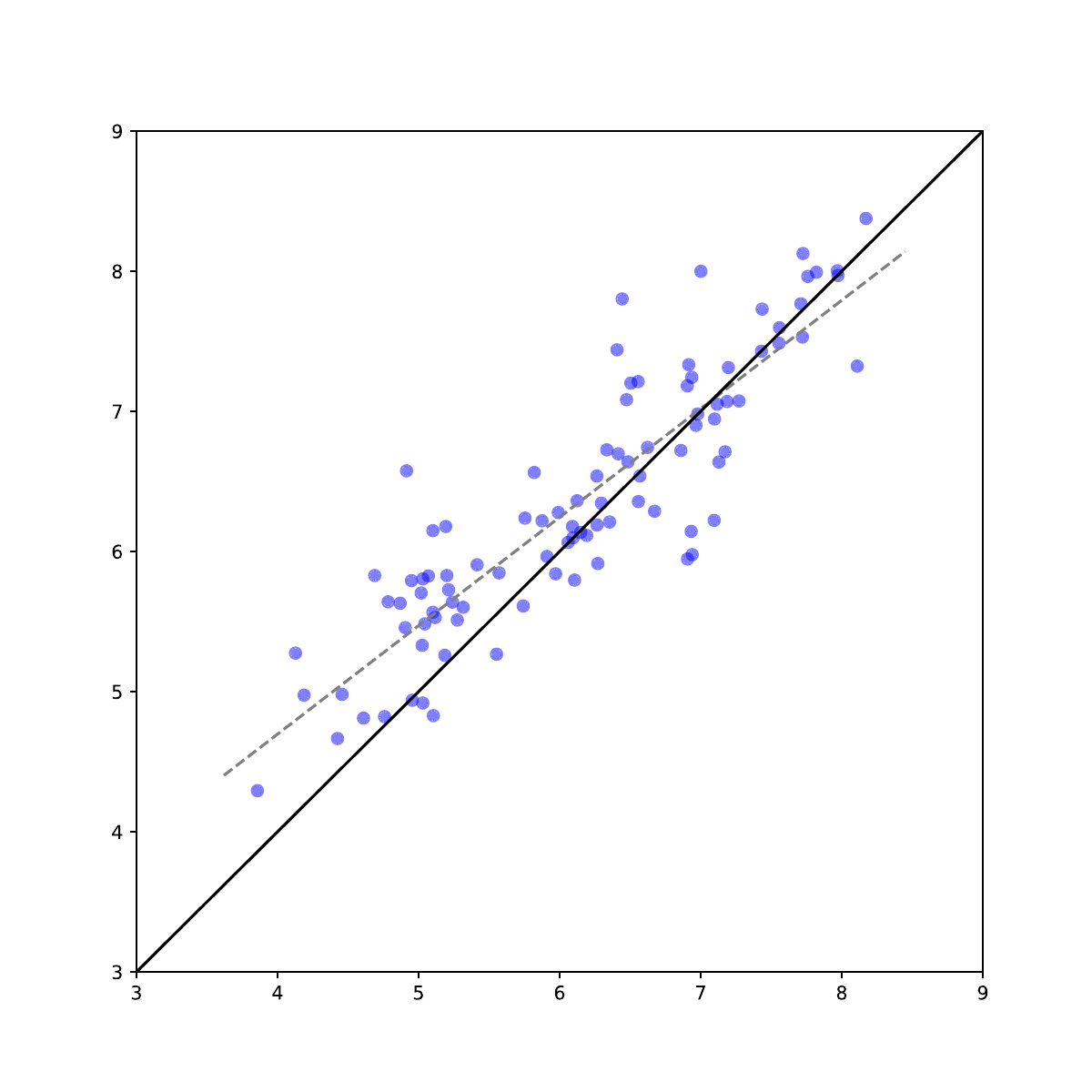}
    \caption{\(R_{nl}, 0.30, 0.79\).}
    \label{fig:2}
\end{subfigure}
\begin{subfigure}{.24\linewidth}
    \centering
\includegraphics[width=0.99\linewidth]{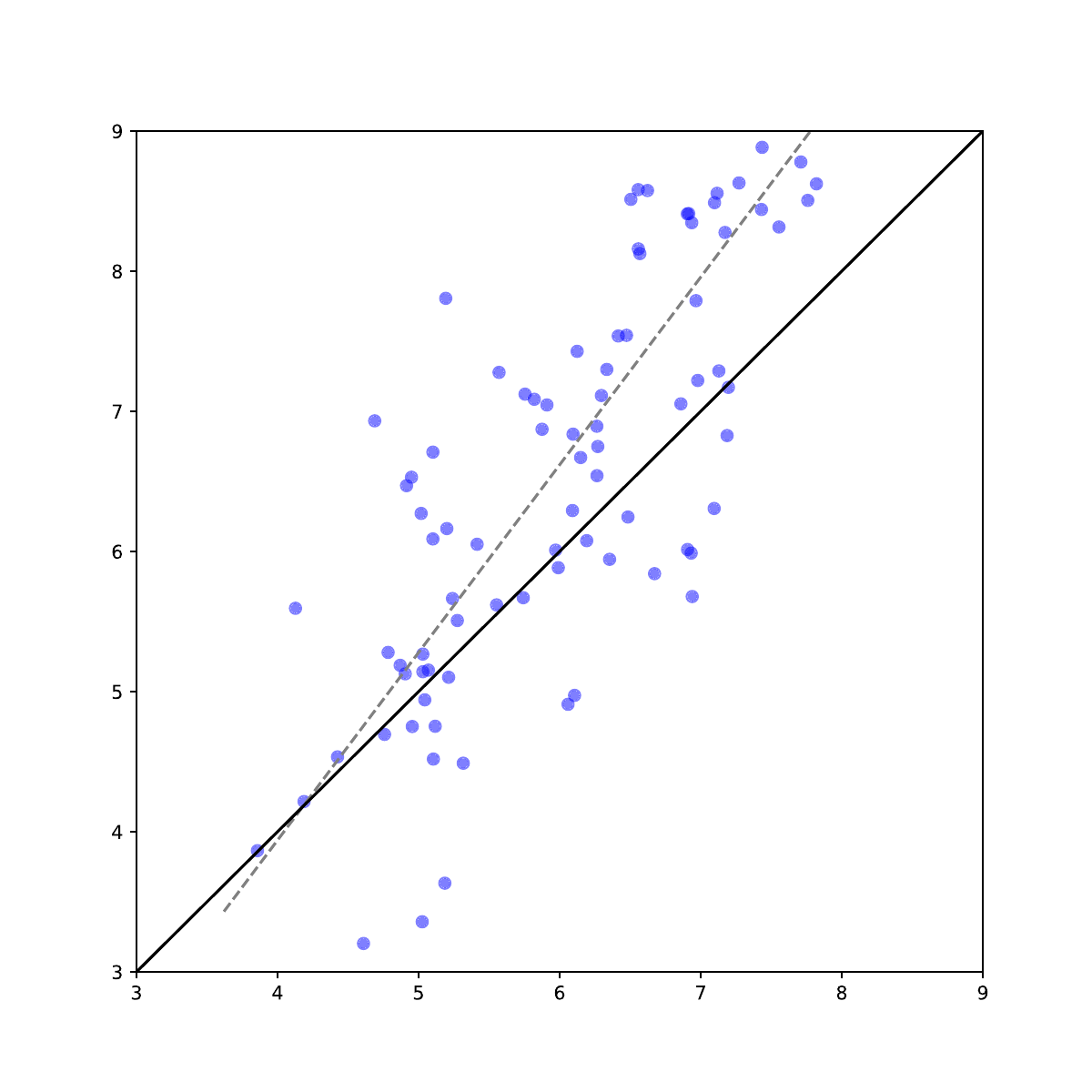}
    \caption{\(B_{l}\, 0.82, 0.66\).}
    \label{fig:5}
\end{subfigure}
\begin{subfigure}{.24\linewidth}
    \centering
    \includegraphics[width=0.99\linewidth]{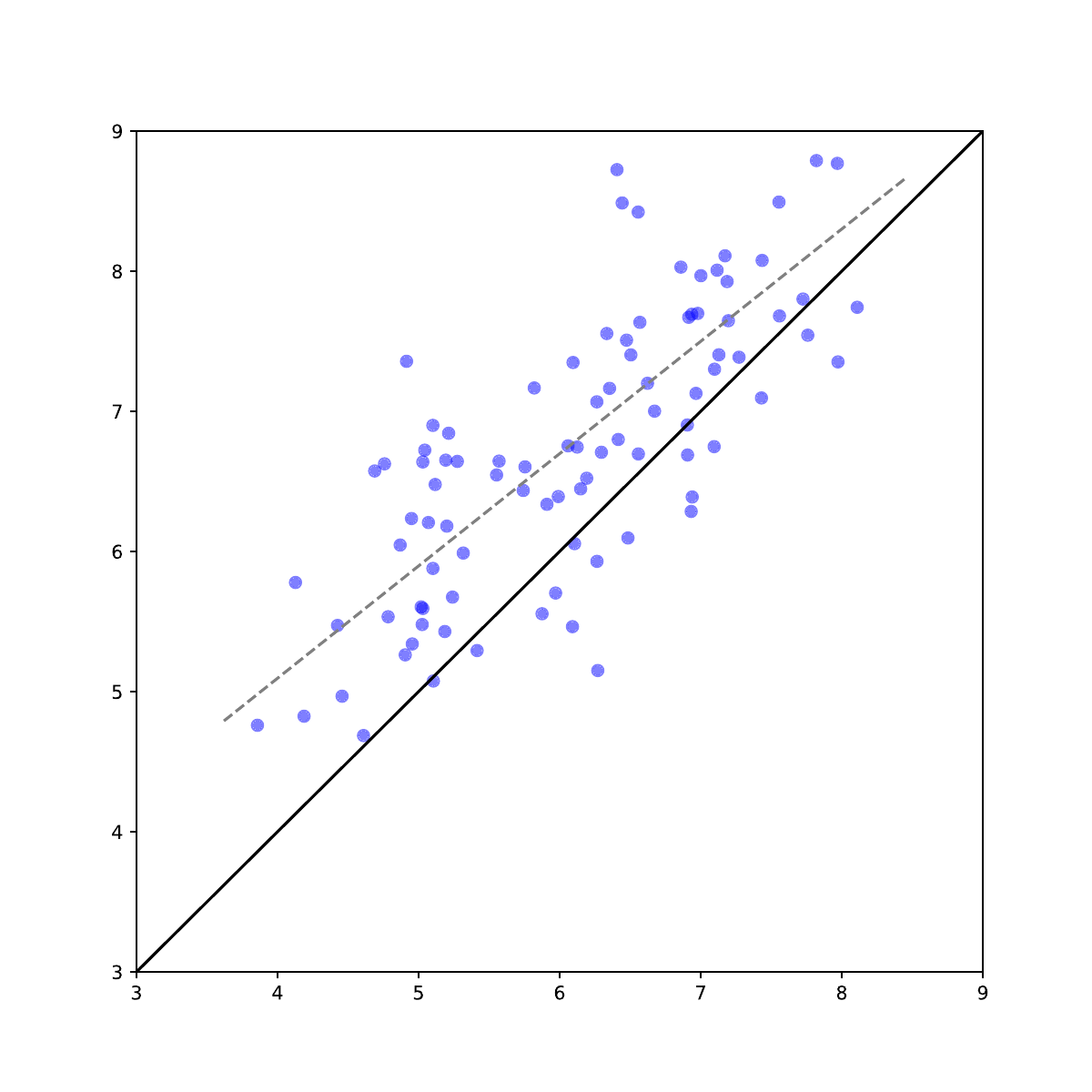}
    \caption{\(B_{nl}, 0.87, 0.6\). }
    \label{fig:6}
\end{subfigure}\caption{Target Variable: NFS}
\end{subfigure}

\caption{A comparison of the four models in question:  \(R_{l},R_{nl},B_{l}\) and $B_{nl}$ using scatter plots. The x-axis represents the actual values, and the y-axis represents the predicted values. Ideally, if the predictions are perfect, the points will lie along a straight solid black line with a slope of 1. Each sub-figure denotes the prediction made by one specific model captioned by its name, Mean Average Error (MEA, the lower - the better), and its determination of coefficients ($R^2$, the higher - the better), respectively. The top row presents the results using the EFS  measure and the bottom row represents the results using the NFS  measure. The gray dashed line indicates the linear regression fitting on the scatter points.} 
\label{fig:model}
\end{figure}

We provide a statistical comparison between $R_l$ and $B_l$, and between $R_{nl}$ and $B_{nl}$, using a standard 80-20 train-test random split. We compare the four model's predictions using the $EFS$ and $NFS$ measures as target variables. 
Fig. \ref{fig:model} depicted the predicted vs actual satisfaction scores. 

Starting with the $EFS$ measure (subfigures (a),(b)), $R_l$ obtained a Mean Absolute Error (MAE) of \(0.70\) with a coefficient of determination \(R^2 = 0.63\) and $R_{nl}$ obtained an MAE of \(0.54\) with a coefficient of determination \(R^2 = 0.74\). In comparison, $B_l$ brings about an MAE of \(1.01\) with \(R^2 = 0.58\)  and $B_{nl}$ brings about an MAE of \(0.84\) with \(R^2 = 0.55\).

Considering the $NFS$ measure,  $R_l$ obtained an MAE of \(0.58\) with a coefficient of determination of \(R^2 = 0.67\) and $R_{nl}$ obtained an MAE of \(0.30\) with a coefficient of determination of \(R^2 = 0.79\). In comparison, $B_l$ brings about an MAE of \(0.82\) with \(R^2 = 0.66\)  and $B_{nl}$ brings about an MAE of \(0.87\) with \(R^2 = 0.60\).

Overall, using both the EFS and NFS measures, the results point to a non-negligible predictive advantage of the $R_l$ and $R_{nl}$ models over the $B_l$ and $B_{nl}$ baseline models, respectively. 

\section{Discussion and Conclusions}
\label{sec:discussion}
In this work, we proposed and evaluated a novel technique for exploring the relationship between the topology of a family tree graph and its members' satisfaction with one another. As illustrated in Fig. \ref{fig:model}, the proposed technique was shown to bring about a noteworthy high accuracy in predicting family members' satisfaction with one another using two common satisfaction measures (EFS and NFS). Furthermore, the results point to the substantial advantage of our technique over using regression models based on key features established in prior literature for the same task. Taken jointly, these combine to suggest that the topology of a family tree graph is, by itself, highly indicative of its members' satisfaction with one another. Moreover, it seems that the connection between the topology of the family tree graph and its members' satisfaction with one another is not straightforward, and cannot be fairly explained through established features from the literature. In particular, the predictive accuracy of a regression model using the nuclear family size, male-to-female ratio, generation age differences, step-siblings and divorce, and oldest generation members' average satisfaction is highly limited in comparison.  
 
We believe that the proposed technique provides a solid foundation for future work and applications. Specifically, future studies exploring novel family characteristics and/or satisfaction measures can use our implementation and results to benchmark their predictive results against a high-performing model that relies solely on the family tree graph's topology. Similarly, other works such as cross-cultural investigations or temporal analyses of family satisfaction dynamics, can use the proposed technique and implementation as an instrument. From an application perspective, our technique could contribute to the development of more informed intervention policies that, in turn, could bring about more efficient resource management policies. Furthermore, given the multiplex nature of the underlying prediction process, it could be valuable to explore the explainability of its outcomes from both theoretical and practical points of view. 

It is important to note that there are several limitations to this work. First, as discussed before, the predictions made using our technique, albeit fairly accurate, are not self-explanatory. As such, we plan to explore explanatory variables through additional, possibly complimentary, tools such as symbolic regression \cite{sym_1,sym_3,sym_2}. If successful, these complimentary new results could provide a stronger theoretical basis for emerging patterns and associations. Second, since our technique relies solely on the family tree graph's topology, we plan to extend it with additional features that are typically not native to the family tree graph such as income, education, and cultural background. Such extension could further increase the predictive ability of our technique and provide new insights into the intricate interaction between these multiplex characteristics. Third, our work considers only \textit{subjective} family satisfaction measures, which may be highly susceptible to various biases (e.g., social desirability bias \cite{new_limit_1}). While this limitation is not unique to our work, future studies could re-examine the issue at hand, using our technique or other means, by using more direct measures of satisfaction such as through bio-markers \cite{new_limit_2}, physiological responses \cite{new_limit_3}, or behavioral observations \cite{new_limit_4}, facilitating deeper, more nuanced and more objective understanding of family members' emotional states and well-being. Finally, focusing on the data used in our evaluation, it is also important to note our sample may not be representative of the entire population. Specifically, we targeted families with good English and satisfactory technological savviness. Naturally, these families may differ in their satisfaction dynamics from other families. In future work, we plan to expand our empirical evaluation to include additional families from varying backgrounds, including those from low socio-economic statuses and/or from marginalized groups, to provide a more comprehensive evaluation.

\section*{Declarations}
\subsection*{Funding}
This research did not receive any specific grant from funding agencies in the public, commercial, or not-for-profit sectors.

\subsection*{Conflicts of interest/Competing interests}

\subsection*{Ethical statement}
This study obtained Ethics approval from the Ariel University ethics community. 

\subsection*{Written informed consent}
All participants of this study agreed to be included using an online written informed consent form.

\subsection*{Data availability}
The data that has been used in this study is available by a formal request from the authors. 

\subsection*{Acknowledgement}
The authors wish to thank Svetlana Hardak-Nisan for inspiring this research. 
 
\bibliography{biblio}
\bibliographystyle{apalike}

\appendix
\section*{Appendix}
Table \ref{table:hyperparamters} presents the hyperparameter values used for the proposed model. The hyperparameters are divided into the encoding part, associated with the VGAE model, and the regression part, associated withe the TPOT model. 

\begin{table}[!ht]
\centering
\begin{tabular}{p{0.5\textwidth}p{0.1\textwidth}}
\hline \hline
\textbf{Hyperparamter} & \textbf{Value} \\ \hline \hline
Number of epochs of the VGAE model & 100 \\
The VGAE model's learning rate  & 0.01 \\
The VGAE model's decay rate & 0.001 \\
The VGAE model's batch size & 8 \\
TPOT's number of folds of the cross-validation (\(k\)) & 5 \\
TPOT's number of generations & 100 \\
TPOT's population size & 100 \\
TPOT's mutation rate & 0.05 \\  \hline \hline
\end{tabular}
\caption{Summary of the hyperparamters and their values.}
\label{table:hyperparamters}
\end{table}

\end{document}